\documentclass[aps,prl,twocolumn,
tightenlines,superscriptaddress,preprintnumbers]{revtex4}

\def\lsim{\raise0.3ex\hbox{$<$\kern-0.75em\raise-1.1ex\hbox{$\sim$}}}
\def\gsim{\raise0.3ex\hbox{$>$\kern-0.75em\raise-1.1ex\hbox{$\sim$}}}

\usepackage{graphicx}

\begin{document}

\title{Shear Viscosity in a Gluon Gas}

\author{Zhe Xu}
\author{Carsten Greiner}
\affiliation{Institut f\"ur Theoretische Physik, Johann Wolfgang
Goethe-Universit\"at Frankfurt, Max-von-Laue-Str.1,
D-60438 Frankfurt am Main, Germany}
\date{\today}

\begin{abstract}
The relation of the shear viscosity coefficient to the recently introduced
transport rate is derived within relativistic kinetic theory. We calculate
the shear viscosity over entropy ratio $\eta/s$ for a gluon gas, which
involves elastic $gg\to gg$ perturbative QCD (PQCD) scatterings as well as
inelastic $gg\leftrightarrow ggg$ PQCD bremsstrahlung. For $\alpha_s=0.3$
we find $\eta/s=0.13$ and for $\alpha_s=0.6$, $\eta/s=0.076$. The small
$\eta/s$ values, which suggest strongly coupled systems, are due to the
gluon bremsstrahlung incorporated.
\end{abstract}

\maketitle

The elliptic flow measurements at the Relativistic Heavy Ion Collider
(RHIC) indicate that the new matter created is a nearly perfect quark
gluon plasma fluid \cite{rhicv2}. Quarks and gluons should be strongly
coupled \cite{sqgp}. The reason for it is still open. Attempts to
understand the phenomena by using perturbative QCD (PQCD), which include
elastic $gg\to gg$ interactions, failed since $gg\to gg$ interactions
cannot even drive the system toward thermalization \cite{SS01}. Also,
the shear viscosity to entropy ratio for elastic processes \cite{AMY00,HG06}
is much larger than the lower bound $\eta/s=1/4\pi$ found from strongly
coupled supersymmetric Yang-Mills gauge theory using the AdS/CFT 
conjecture \cite{adscft}. However, the importance of PQCD bremsstrahlung
is raised in the ``bottom-up'' thermalization scenario \cite{BMSS01} and
the physics regarding jet-quenching \cite{jet}. Recent calculations within
the Boltzmann approach of multiparton scatterings (BAMPS) \cite{XG05,XG07},
which includes PQCD $gg\leftrightarrow ggg$ bremsstrahlung, demonstrated that
the $gg\leftrightarrow ggg$ processes are a factor of $5$ more efficient
for thermal equilibration and pressure buildup
than the $gg\to gg$ scatterings, whereas their transition rates are almost
the same. It raises again the question whether PQCD interactions
can in fact explain that the quark gluon plasma behaves like
a ``strongly coupled'' system with a small shear viscosity to entropy ratio.

In this Letter we first derive a useful formula for the shear viscosity
coefficient within relativistic kinetic theory and then
calculate the shear viscosity to entropy ratio for a gluon gas in thermal
equilibrium, which includes PQCD $gg\to gg$ and
$gg\leftrightarrow ggg$ interactions.

For a gas the phase space distribution of particles $f(p,x)$
satisfies the Boltzmann equation
\begin{equation}
\label{boltzmann}
v^{\mu}\partial_{\mu} f=I\,,
\end{equation}
where $v^{\mu}=(1,{\bf p}/E)$ and $I$ denotes the collision term that
determines the change in $f$ due to interactions among particles.
If the gas is locally in kinetic equilibrium but still away from chemical 
equilibrium, its phase space distribution is described by the distribution
\begin{equation}
\label{feq}
f_0=\left ( e^{\beta u_{\mu}p^{\mu}}/\lambda \mp 1 \right )^{-1}\,,
\end{equation}
where $\lambda(x)$ is the fugacity, $\beta(x)$ is the inverse of temperature
$T(x)$, and $u_{\mu}(x)$ denotes the four-velocity of the medium.
The $\mp$ sign applies to bosons and fermions, respectively.

The viscosity can be extracted from the response of the medium to
a small deviation from $f_0$
\begin{equation}
\label{f1}
f=f_0+f_0(1\pm f_0)f_1
\end{equation}
by comparing the stress tensor defined as
\begin{equation}
T^{\mu \nu}=\int \frac{d^3p}{(2\pi)^3 E} p^{\mu} p^{\nu} f
\end{equation}
with its expression in the Navier-Stokes approximation
\begin{equation}
\label{ns1}
T^{\mu \nu}=(\epsilon + P) u^{\mu} u^{\nu} -Pg^{\mu \nu}+\tau^{\mu \nu}\,,
\end{equation}
where
\begin{equation}
\label{ns2}
\tau^{\mu \nu}=\eta \, (\nabla^{\mu}u^{\nu}+\nabla^{\nu}u^{\mu}
-\frac{2}{3}\Delta^{\mu \nu}\nabla_{\alpha}u^{\alpha})+
\xi\, \Delta^{\mu \nu}\nabla_{\alpha}u^{\alpha}\,.
\end{equation}
$\eta$ denotes the shear viscosity and $\xi$ the bulk viscosity,
and $\Delta^{\alpha \beta}=g^{\alpha \beta}-u^{\alpha}u^{\beta}$,
$\nabla^{\alpha}=\Delta^{\alpha \beta} \partial_{\beta}$ where
$g^{\mu \nu}={\rm diag}(1,-1,-1,-1)$. For (\ref{ns2})
Landau's definition of $u^{\mu}$ is used \cite{degroot}. 
In the rest frame of the medium we obtain
\begin{equation}
\label{shv}
\eta=\frac{T_{xx}+T_{yy}-2\,T_{zz}}{2\,(3\,\partial_z u_z
-\vec \nabla\cdot \vec u)}\,,
\end{equation}
\begin{equation}
\label{bv}
\xi=\frac{3\,P-T_{xx}-T_{yy}-T_{zz}}{3\,\vec \nabla\cdot \vec u}\,.
\end{equation}
In a conformal theory, which can be applied to the (massless) gluon gas
at high temperature, $T_{xx}+T_{yy}+T_{zz}=T_{00}=\epsilon=3P$ so that
the bulk viscosity vanishes.

With (\ref{f1}) and using
\begin{equation}
\label{df0}
\partial_{\mu} f_0=f_0(1\pm f_0)\left [ \partial_{\mu} \ln \lambda
-\partial_{\mu}(\beta u_{\nu}p^{\nu}) \right ]\,,
\end{equation}
the left-hand side of the Boltzmann equation (\ref{boltzmann}) becomes
\begin{eqnarray}
\label{df}
v^{\mu} \partial_{\mu} f&=& f_0(1\pm f_0) \left [1+(1\pm f_0)f_1\pm f_0f_1
\right ] \nonumber\\
&& \times v^{\mu} \left [ \partial_{\mu} \ln \lambda
-\partial_{\mu}(\beta u_{\nu}p^{\nu}) \right ]\nonumber \\
&&+f_0(1\pm f_0) v^{\mu} \partial_{\mu} f_1\nonumber \\
&\cong& f v^{\mu} \partial_{\mu} \ln \lambda
-f_0(1\pm f_0)v^{\mu}\partial_{\mu}(\beta u_{\nu}p^{\nu})\,,
\end{eqnarray}
where the neglected terms are of second order in spatial gradients when
assuming that $f_1$ in (\ref{f1}) only contains terms of first order
in spatial gradients. Following $\partial_{\mu} T^{\mu \nu}=0$,
Eq. (\ref{df}) is rewritten in the rest frame
\begin{eqnarray}
\label{df2}
v^{\mu} \partial_{\mu} f \cong && f v^{\mu} \partial_{\mu} \ln \lambda
+f_0(1\pm f_0)\frac{\partial \beta}{\partial \epsilon} \nonumber \\
&& \times \left [ \partial_t(\ln\lambda) E \epsilon+3\partial_{\mu}(\ln\lambda)
P^i T^{\mu i} \right ] \nonumber \\
&& + f_0(1\pm f_0)\beta E v^i v^j \nonumber \\
&& \times \left [ \frac{1}{2} (\partial_iu^j+
\partial_ju^i)-\frac{1}{3} \delta_{ij} \vec \nabla \cdot \vec u \right ]\,,
\end{eqnarray}
where $i,j=1,2,3$ and the equation of state $\epsilon=3P$ is used for 
massless particles. For $\lambda=1$ Eq. (\ref{df2}) is identical with
the term derived by Arnold {\it et al.} in \cite{AMY00}.

Integrating Eq. (\ref{boltzmann}) over momentum using (\ref{df2}) gives
\begin{equation}
\label{mom0}
\int \frac{d^3p}{(2\pi)^3} I=\int \frac{d^3p}{(2\pi)^3}
v^{\mu} \partial_{\mu} f \cong
\frac{1}{4} n \partial_t (\ln \lambda)\,,
\end{equation}
where 
\begin{equation}
\label{dens}
n=\int \frac{d^3p}{(2\pi)^3} f
\end{equation}
is the particle density. In the derivation for (\ref{mom0}) we assumed
that there is no particle flow in the rest frame following
Eckard's definition of $u^{\mu}$ \cite{degroot}.
Integrating (\ref{boltzmann}) by weight $v_z^2$ gives
\begin{eqnarray}
\label{mom2}
&&\int \frac{d^3p}{(2\pi)^3} v_z^2 I = \int \frac{d^3p}{(2\pi)^3}
v_z^2 v^{\mu} \partial_{\mu} f \nonumber \\
&\cong& \frac{2}{15} n \left (3 \, \partial_z u_z -\vec \nabla \cdot \vec u
\right ) + \left (\frac{1}{4}-\langle v_z^2 \rangle \right ) n \partial_t
(\ln \lambda)
\end{eqnarray}
where $\langle \cdot \rangle$ denotes the average over particles.
We then obtain the relation
\begin{equation}
\label{vg}
3 \, \partial_z u_z -\vec \nabla \cdot \vec u
\cong \frac{15}{2} \left (\frac{1}{3}-\langle v_z^2 \rangle \right )
\left ( \sum R^{\rm tr}+\frac{3}{4} n \partial_t (\ln \lambda)
\right )\,,
\end{equation}
where 
\begin{equation}
\label{trate}
\sum R^{\rm tr}= \frac{\int \frac{d^3p}{(2\pi)^3} v_z^2 I -
\langle v_z^2 \rangle \int \frac{d^3p}{(2\pi)^3} I}{n\, (\frac{1}{3}-
\langle v_z^2 \rangle)}
\end{equation}
is the total transport collision rate, which was introduced in \cite{XG07}
as the characteristic quantity describing momentum isotropization. In
kinetic equilibrium, Eq. (\ref{vg}) becomes an exact equation.

Inserting (\ref{vg}) into (\ref{shv}) we obtain
\begin{equation}
\label{shv2}
\eta \cong \frac{1}{5} n \frac{\langle E(\frac{1}{3}-v_z^2) \rangle}
{\frac{1}{3}-\langle v_z^2 \rangle} \frac{1}{\sum R^{\rm tr}+ 
\frac{3}{4} n \partial_t (\ln \lambda)}\,.
\end{equation}
This expression constitutes our major formula calculating the shear viscosity
coefficient $\eta$ and gives a direct correspondence of $\eta$
to the transport rate $R^{\rm tr}$: $\eta$ is inversely
proportional to the sum of the total transport collision rate
and the chemical equilibration rate, and is roughly proportional to the
energy density. If the chemical equilibration
is governed by $2\leftrightarrow 3$ processes,
\begin{equation}
\label{chem}
\frac{3}{4} n \partial_t (\ln \lambda)= 3 \int \frac{d^3p}{(2\pi)^3} I
=\frac{3}{2}R_{23}-R_{32}\,,
\end{equation}
which might even become negative for systems with oversaturation. 
However, if the system is not far away from chemical equilibrium, the total
transport collision rate is most dominant and, thus, determines the shear
viscosity.

If $f$ or $f_1$ in (\ref{f1}) is known and $f_1 \ne 0$, $\eta$ can be in
principle calculated using (\ref{shv}). A way to get $f_1$ is to solve
the linearized Boltzmann equation
\begin{equation}
\label{getf1}
f v^{\mu} \partial_{\mu} \ln \lambda
-f_0(1\pm f_0)v^{\mu}\partial_{\mu}(\beta u_{\nu}p^{\nu})= I(f_1)
\end{equation}
as a variational problem for $f_1$ \cite{degroot}. Often used is a
simple ansatz \cite{AMY00,ABM07} for $f_1$ such as the function
(\ref{fform2}) below. One notices that $\eta$ obtained from (\ref{shv})
using $f$ from (\ref{getf1}) is identical with that from (\ref{shv2}).
Another more complicated method to get $f$ is to solve the Boltzmann
equation (\ref{boltzmann}) numerically performing extensive transport 
simulations, which is, in principle, more reliable than the first method
except for numerical uncertainties. Since the spatial gradients are needed in
(\ref{shv}) and their extractions from transport simulations are difficult,
it is at present more convenient to use (\ref{shv2}) to calculate $\eta$.
Such calculations will be presented in a forthcoming publication.

In this Letter we consider a static particle system, which is initially
disturbed from equilibrium and is relaxing again to thermal equilibrium.
For a Boltzmann gas in chemical equilibrium $f$ is assumed to have
the form \cite{RS03}
\begin{equation}
\label{fform1}
f=e^{-\beta \sqrt{E^2+\chi p_z^2}}
\end{equation}
in the rest frame. For small $\chi$ 
\begin{equation}
\label{fform2}
f\approx e^{-\beta E} \left (1-\frac{\chi}{2}\, \beta \,\frac{p_z^2}{E}
\right )\,.
\end{equation}
Comparing (\ref{fform2}) with (\ref{f1}) 
$f_1=-\frac{\chi}{2}\, \beta \,\frac{p_z^2}{E}$,
which satisfies Eq. (\ref{getf1}) with appropriate $\chi$.
Using (\ref{fform2}) in (\ref{shv2}) we calculate $\eta$
in thermal equilibrium as the limit when letting $\chi \to 0$.
In this limit Eq. (\ref{shv2}) becomes an exact equation identical to
Eq. (\ref{shv}). We obtain
\begin{equation}
\label{shv3}
\eta=\frac{4}{15} \, \epsilon \, l^{\rm tr}\,,
\end{equation}
where $l^{\rm tr}$ is the inverse of the total transport collision rate
and is called the mean transport path.

We now apply Eq. (\ref{shv3}) to calculate the shear viscosity in a gluon
gas. For the sake of simplicity we assume gluons to be Boltzmann particles.
The boson enhancement is neglected, which is a good approximation for
gluons at high temperatures. Interactions among gluons include
elastic $gg\to gg$ PQCD scatterings and inelastic
$gg\leftrightarrow ggg$ PQCD bremsstrahlung processes, which are screened
by the Debye mass
\begin{equation}
\label{md}
m_D^2=\pi \, d_G \,\alpha_s \int \frac{d^3p}{(2\pi)^3} \frac{1}{p} N_c f_0
=\frac{1}{2\pi} \, d_G \, N_c \, \alpha_s \frac{1}{\beta^2}\,,
\end{equation}
where $d_G=16$ denotes the gluon degeneracy factor and $N_c=3$.
The Landau-Pomeranchuk-Migdal suppression of bremsstrahlung is taken
into account as a lower cutoff in the momentum of the radiated
gluon \cite{B93}.
The matrix elements of the transitions can be found in \cite{XG05,XG07}.
We obtain $\sum R^{\rm tr}$ in thermal equilibrium as the term in $0$th order
of $\chi$ in Eq. (\ref{trate}) when using Eq. (\ref{fform2}) for the integrals.
Because the collision term is additive,
$\sum R^{\rm tr}=R^{\rm tr}_{gg \to gg}+R^{\rm tr}_{gg \to ggg}
+R^{\rm tr}_{ggg \to gg}$
where $R^{\rm tr}_{ggg \to gg}=\frac{2}{3}R^{\rm tr}_{gg \to ggg}$ due to
detailed balance \cite{XG07}.

\begin{figure}[b]
\begin{center}
\includegraphics[height=8.5cm]{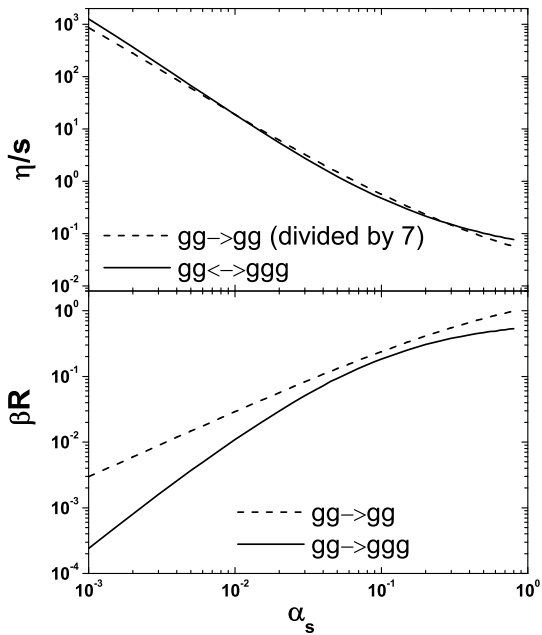}
\end{center}
\vspace{-0.5cm}
\caption{Upper panel: Shear viscosity to entropy ratio for $gg\to gg$ and 
$gg \leftrightarrow ggg$ processes. $\eta/s$ for $gg\to gg$ is divided
by a factor of $7$. Lower panel: Collision rate to temperature ratio for
$gg\to gg$ and $gg \to ggg$ processes.
}
\label{fig:etas}
\end{figure}

The shear viscosity to entropy ratio at equilibrium is
\begin{equation}
\label{etas}
\frac{\eta}{s}=\left ( 5 \beta \sum R^{\rm tr} \right )^{-1}=
\left (5 \beta R^{\rm tr}_{gg \to gg}+ 
\frac{25}{3} \beta R^{\rm tr}_{gg \to ggg} \right )^{-1}\,,
\end{equation}
where $s=\frac{4}{3} \beta \epsilon$ is used. Because the transport collision
rates scale with the temperature, $\beta R^{\rm tr}_{gg\to gg}$,
$\beta R^{\rm tr}_{gg\to ggg}$ and, thus, $\eta/s$
depend only on the coupling constant $\alpha_s$. The upper panel of 
Fig. \ref{fig:etas} shows the shear viscosity to entropy ratio for
$gg\to gg$ and $gg \leftrightarrow ggg$ processes, respectively.
$\eta/s$ for $gg \leftrightarrow ggg$ processes is roughly a factor of $7$
smaller than that for the elastic collisions, which implies that compared to
the elastic scatterings the PQCD inspired bremsstrahlung is the leading
process in relaxing the system to equilibrium.
For $\alpha_s=0.3$, which might be appropriate at RHIC energy, 
$\eta/s=1.03$ for $gg\to gg$ only and $\eta/s=0.13$ when including 
$gg \leftrightarrow ggg$ processes. To match the lower bound of
$\eta/s=1/4\pi$ from the AdS/CFT conjecture \cite{adscft} $\alpha_s=0.6$
has to be chosen.
The $\eta/s$ ratios in the two cases correspond to the mean transport path 
$l^{\rm tr}=1/\sum R^{\rm tr}=0.32$ fm and $l^{\rm tr}=0.2$ fm at
$T=1/\beta=400$ MeV for $\alpha_s=0.3$ and $\alpha_s=0.6$, respectively.
From the collision rates shown in the lower panel of
Fig. \ref{fig:etas} with $n=d_G\,T^3/\pi^2$ we obtain 
$\langle v_{\rm rel} \sigma_{gg\to gg} \rangle=R_{gg\to gg}/n=0.82$ mb and
$\langle v_{\rm rel} \sigma_{gg\to ggg} \rangle=R_{gg\to ggg}/n=0.57$ mb
for $\alpha_s=0.3$, and 
$\langle v_{\rm rel} \sigma_{gg\to gg} \rangle=1.27$ mb and
$\langle v_{\rm rel} \sigma_{gg\to ggg} \rangle=0.73$ mb for
$\alpha_s=0.6$ at $T=400$ MeV. 
Hence, perturbative interactions can drive gluons to behave like
a strongly coupled system with a small $\eta/s$ ratio at RHIC.

\begin{figure}[b]
\begin{center}
\includegraphics[height=8.5cm]{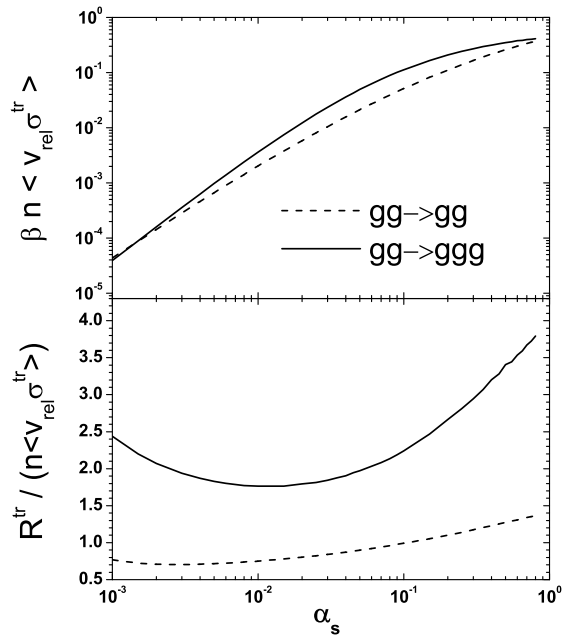}
\end{center}
\vspace{-0.5cm}
\caption{Scaled transport cross section (upper panel) and transport collision
rate to scaled transport cross section ratio (lower panel) 
for $gg\to gg$ and $gg \to ggg$ processes.
}
\label{fig:trcs}
\end{figure}

From Fig. \ref{fig:etas} we also obtain that 
$R^{\rm tr}_{gg\to gg}/R_{gg\to gg}=0.36(0.46)$ and 
$R^{\rm tr}_{gg\to ggg}/R_{gg\to ggg}=2.1(2.7)$ for $\alpha_s=0.3(0.6)$,
respectively. The wide difference in the behavior of the
$R^{\rm tr}/R$ ratio for the $gg \to gg$ and $gg\to ggg$ processes
is essential for the different contributions to $\eta/s$.
Because $R^{\rm tr}$ contains an indirect relationship with the distribution
of the collision angle $\theta$, we decompose the transport collision rate to
\begin{equation}
\label{sepa}
R^{\rm tr}_i= A_i \, n \langle v_{\rm rel} \sigma^{\rm tr}_i\rangle
\,, \quad i=gg\to gg\,,\ gg \to ggg\,,
\end{equation}
with $\sigma^{\rm tr}_i=\int d\sigma_i \sin^2\theta$ defined as the transport
cross section \cite{DG85} and $A_i$ being a multiplication factor.
Figure \ref{fig:trcs} shows 
$\beta n \langle v_{\rm rel} \sigma^{\rm tr}_i \rangle$ and $A_i$ as
function of $\alpha_s$.
For $\alpha_s=0.3(0.6)$ 
$\langle v_{\rm rel} \sigma^{\rm tr}_{gg\to gg} \rangle /
\langle v_{\rm rel} \sigma_{gg\to gg} \rangle=0.3(0.35)$ and 
$\langle v_{\rm rel} \sigma^{\rm tr}_{gg\to ggg} \rangle /
\langle v_{\rm rel} \sigma_{gg\to ggg} \rangle=0.71(0.78)$, which
indicate that for the chosen $\alpha_s$ values $gg\to gg$ favors
small-angle scattering and $gg\to ggg$ favors large-angle 
bremsstrahlung \cite{XG07}.

The factors $A_i$ in Eq. (\ref{sepa}) have weak dependences on
$\alpha_s$ and are around the isotropic distribution values.
(For isotropic angular distribution
$A_{gg\to gg}=9/8$ and $A_{gg\to ggg}=27/16$.) For $\alpha_s=0.3(0.6)$
we obtain $A_{gg\to ggg}/A_{gg\to gg}=2.5(2.7)$, which are significantly
larger than $1$.

Parametrically, $\beta R^{\rm tr}_{gg \to gg}$ is fitted by
$(0.68+2.8\alpha_s)\, \alpha_s^2 \, (\ln \alpha_s)^2$
from $\alpha_s=0.001$ up to $0.3$, whereas 
$\beta R_{gg\to gg} \approx (3-36\alpha_s/\pi)\alpha_s$ for
$\alpha_s < 0.1$. Thus, 
$R^{\rm tr}_{gg \to gg}/R_{gg \to gg} \sim \mathcal{O}(\alpha_s)$
for $\alpha_s < 0.1$, which indicates again small-angle $gg\to gg$
scatterings, because for those scatterings
$R^{\rm tr}_{gg \to gg}/R_{gg \to gg} \sim \theta^2$ with
$\theta^2 \approx 4q^2/s_m \approx 4m_D^2/s_m \sim \alpha_s$ where
$q$ is the momentum transfer.

Including the Landau-Pomeranchuk-Migdal effect for the PQCD inspired
bremsstrahlung within the Bethe-Heitler regime 
$\beta R_{gg\to ggg}\sim \alpha_s^2 (\ln \alpha_s)^2$ \cite{XG05}.
For small-angle bremsstrahlung $\beta R^{\rm tr}_{gg\to ggg}$ should be
of the order $\alpha_s^3 (\ln \alpha_s)^2$, which is clearly not the case
for $\alpha_s > 0.01$ as seen from Fig. \ref{fig:trcs} as within this
$\alpha_s$ range the transport cross sections for $gg\to gg$ and
$gg\to ggg$ have almost the same $\alpha_s$ dependence, namely 
$\mathcal{O}[\alpha_s^2 \, (\ln \alpha_s)^2]$. Thus, 
$R^{\rm tr}_{gg \to ggg}/R_{gg \to ggg} \sim \mathcal{O}(1)$ for 
$\alpha_s > 0.01$, which shows that in the chosen $\alpha_s$
interval the bremsstrahlung favors isotropic angular distribution.
For smaller $\alpha_s$, however, the collision angle tends to be
distributed forward. This can be observed in Fig. \ref{fig:trcs} where
$\beta n \langle v_{\rm rel} \sigma^{\rm tr}_{gg\to ggg} \rangle$
becomes steeper than 
$\beta n \langle v_{\rm rel} \sigma^{\rm tr}_{gg\to gg} \rangle$
at $\alpha_s < 0.01$. This implies that the ``collinear'' bremsstrahlung where
$R^{\rm tr}_{gg \to ggg}/R_{gg \to ggg} \sim \mathcal{O}(\alpha_s)$
will occur at extreme small $\alpha_s$ and only then will have smaller
contribution to the transport coefficients than the $gg\to gg$
processes \cite{AMY03}.

Finally, the ratio of the collisional width to the mean gluon energy,
$\Gamma/\langle E \rangle=(R_{gg\to gg}+R_{gg\to ggg}+R_{ggg\to gg})\beta/3$,
can be calculated from the lower panel of Fig. \ref{fig:etas} where 
$R_{ggg\to gg}=1.5 R_{gg\to ggg}$ due to detailed balance.
For $\alpha_s=0.3(0.6)$ we obtain $\Gamma/\langle E \rangle=0.5(0.69)$.
These ratios are smaller than, but, close to $1$, which indicates that
PQCD with $\alpha_s=0.3-0.6$ is at the edge of its applicability. 
For larger values of $\alpha_s$ one obtains $\Gamma \ge \langle E \rangle$
and the PQCD calculations with on shell kinematics are no longer applicable,
as for such a strong coupling regime the Heisenberg uncertainty principle
has to be taken care of by a full quantum transport treatment \cite{JCG04}.

The higher order processes such as $ggg\to ggg$ and
$gg\leftrightarrow gggg$ will certainly modify the total transition
rate. However, their contributions are suppressed by higher order of 
$\alpha_s$ \cite{XS94}. Because the full diagrammatic many-body theory
for higher order collisions becomes rather complex \cite{BD68}, at present,
the incorporation of higher order multiparticle interactions is considered
in a phenomenological manner \cite{D90}.


In summary, the shear viscosity is derived
within relativistic kinetic theory, which is proportional
to energy density and inversely proportional to the total transport
collision rate. We calculated the shear viscosity to entropy ratio $\eta/s$
for a gluon gas, and found $\eta/s=0.13(0.076)$ for $\alpha_s=0.3(0.6)$.
Perturbative QCD interactions can drive the gluon matter to a strongly
coupled system with an $\eta/s$ ratio as small as the lower bound from
the AdS/CFT conjecture. The PQCD inspired gluon bremsstrahlung is
responsible for small $\eta/s$ ratios and, thus, can explain that
the quark gluon plasma created at RHIC behaves like a nearly perfect fluid.

\acknowledgments
Z.~X. would like to thank N.~Su for fruitful discussions.

\end{document}